\documentclass[conference]{IEEEtran}
\usepackage{cite}
\usepackage{ragged2e}
\usepackage{amsmath,amssymb,amsfonts}
\usepackage{algorithmic}
\usepackage{graphicx}
\usepackage{textcomp}
\usepackage{caption}
\usepackage{fancyhdr}
\usepackage{subcaption}
\usepackage{xcolor}
\usepackage{multirow}
\def\BibTeX{{\rm B\kern-.05em{\sc i\kern-.025em b}\kern-.08em
    T\kern-.1667em\lower.7ex\hbox{E}\kern-.125emX}}

\makeatletter
\newcommand*\titleheader[1]{\gdef\@titleheader{#1}}
\AtBeginDocument{%
	\let\st@red@title\@title
	\def\@title{%
		\bgroup\normalfont\large\raggedright\@titleheader\par\egroup
		\vskip0.5em\st@red@title}
}
\makeatother
\titleheader{This work has been submitted to the IEEE for possible publication. Copyright may be transferred without notice, after which this version may no longer be accessible.}

\title{Classification of COVID-19 in Chest X-ray Images Using Fusion of Deep Features and LightGBM}

\makeatletter
\newcommand{\linebreakand}{%
  \end{@IEEEauthorhalign}
  \hfill\mbox{}\par
  \mbox{}\hfill\begin{@IEEEauthorhalign}
}
 \makeatother

\author{\IEEEauthorblockN{\hspace{0.5cm}Hamid Nasiri}
 \IEEEauthorblockA{\hspace{0.5cm}\textit{Department of Computer Engineering} \\
 \hspace{0.5cm}\textit{Amirkabir University of  Technology}\\
 \hspace{0.5cm}Tehran, Iran \\
\hspace{0.5cm}h.nasiri@aut.ac.ir}
\and
\IEEEauthorblockN{\hspace{1cm}Ghazal Kheyroddin}
 \IEEEauthorblockA{\hspace{1cm}\textit{Electrical and Computer Engineering Department} \\
 \hspace{1cm}\textit{Semnan University}\\
 \hspace{1cm}Semnan, Iran \\
\hspace{1cm}gh.kheyroddin@gmail.com}
\linebreakand
 \IEEEauthorblockN{\hspace{-0.2cm}Morteza Dorrigiv}
\IEEEauthorblockA{\hspace{-0.2cm}\textit{Electrical and Computer Engineering Department} \\
 \hspace{-0.2cm}\textit{Semnan University}\\
 \hspace{-0.2cm}Semnan, Iran \\
\hspace{-0.2cm}dorrigiv@semnan.ac.ir}
 \and
\IEEEauthorblockN{Mona Esmaeili}
\IEEEauthorblockA{\textit{Electrical and Computer Engineering Department} \\
 \textit{University of New Mexico}\\
 Albuquerque, USA \\
 mesmaeili@unm.edu}
 \linebreakand
 \IEEEauthorblockN{\hspace{-0.8cm}Amir Raeisi Nafchi}
 \IEEEauthorblockA{\hspace{-0.8cm}\textit{Electrical and Computer Engineering Department} \\
 \hspace{-0.8cm}\textit{University of New Mexico}\\
\hspace{-0.8cm}Albuquerque, USA \\
\hspace{-0.8cm}amirorn@unm.edu}
\and
\IEEEauthorblockN{Mohsen Haji Ghorbani}
\IEEEauthorblockA{\textit{Department of Computer Engineering} \\
\textit{Technical and Vocational University (TVU)}\\
Tehran, Iran \\
 mohsen.hajighorbani@tvu.ac.ir}
\linebreakand
 \IEEEauthorblockN{Payman Zarkesh-Ha}
 \IEEEauthorblockA{\textit{Electrical and Computer Engineering Department} \\
 \textit{University of New Mexico}\\
Albuquerque, USA \\
pzarkesh@unm.edu}
 }

\begin{document}

\maketitle

\IEEEpubidadjcol
\begin{abstract}
The COVID-19 disease was first discovered in Wuhan, China, and spread quickly worldwide. After the COVID-19 pandemic, many researchers have begun to identify a way to diagnose the COVID-19 using chest X-ray images. The early diagnosis of this disease can significantly impact the treatment process.
In this article, we propose a new technique that is faster and more accurate than the other methods reported in the literature. 
The proposed method uses a combination of DenseNet169 and MobileNet Deep Neural Networks to extract the features of the patient's X-ray images. Using the univariate feature selection algorithm, we refined the features for the most important ones. Then we applied the selected features as input to the LightGBM (Light Gradient Boosting Machine) algorithm for classification. 
To assess the effectiveness of the proposed method, the ChestX-ray8 dataset, which includes 1125 X-ray images of the patient's chest, was used. The proposed method achieved 98.54$\%$ and 91.11$\%$ accuracies in the two-class (COVID-19, Healthy) and multi-class (COVID-19, Healthy, Pneumonia) classification problems, respectively. It is worth mentioning that we have used Gradient-weighted Class Activation Mapping (Grad-CAM) for further analysis.  

\end{abstract}

\IEEEpeerreviewmaketitle
\section{Introduction}

Scientists discovered a family of viruses known as coronaviruses in 1965\cite{mahase2020covid}. These viruses range from the cold virus to the COVID-19 virus. However, seven different human-transmitted coronaviruses have been discovered. The latest one, severe acute respiratory syndrome coronavirus 2 (SARS-CoV-2), was found in Wuhan, China. World Health Organization (WHO) named the disease caused by SARS-CoV-2 COVID-19.
The common symptoms of COVID-19 disease include cough, shortness of breath, sore throat, fever, and body pain \cite{struyf2021signs}.\\ \indent
COVID-19 can cause lung injury and death in severe cases if not treated. By September 2020, more than 223 million people have contacted the COVID-19, and over 4.6 million people have died due to COVID-19.
The best method of diagnosing coronavirus is the RT-PCR test.
Because taking PCR test takes time, doctors use the chest computed tomography scan (CT scan) for diagnosis \cite{singhal2020review}. 
This study uses the patient's chest X-ray images and offers a novel method for early discovery of COVID-19 infection. 

This paper suggests MobileNet and DenseNet169 Deep Neural Networks (DNNs) for feature extraction from the chest X-ray images. After combining the characteristics extracted from both networks, the most significant features are selected and given as input to the Light Gradient Boosting Machine (LightGBM) classification algorithm to classify the images.

\section{Related Works}
Ozturk et.al. \cite{ozturk2020automated} utilized the DarkNet network in 2020 for their research. The training model is made up of 17 convolution layers. They obtained an accuracy of 87.02$\%$ in the three-class classification and 98.8$\%$ in the two-class classification. Their research was based on a dataset containing 1125 images, including 500 lung images from healthy people, 500 lung images from people with Pneumonia, and 125 images from people with COVID-19. Abbas et al. \cite{abbas2021classification} used the DeTraC deep convolutional neural network architecture to classify the two classes, with an accuracy of 93.1$\%$. They used a dataset of images from the chest of 126 people with COVID-19 disease and 80 healthy people to conduct their research. Minaei and his colleagues\cite{minaee2020deep} have used four models with convolutional architecture to address the problem of both classes. They include ResNet18, SqueezeNet, ResNet50 and DenseNet121. 

Nasiri and Alavi \cite{nasiri2022novel} proposed a novel framework based on deep learning and analysis of variance (ANOVA) feature selection method. They used XGBoost for classification.
Gunraj and colleagues \cite{gunraj2020covidnet}, in their research on convolution architecture networks, used COVID-Net and reported 93.3$\%$ accuracy. Hemdan and his colleagues\cite{hemdan2020covidx} obtained an accuracy of 90$\%$, after examining different networks using the VGG19 and DenseNet201 networks. Narin\cite{narin2021automatic} and colleagues selected the ResNet50 neural network by testing five pre-trained neural networks on three separate datasets, using this neural network to achieve an average accuracy of 98$\%$ in a two-class problem. Barstugan\cite{barstuugan2020coronavirus} extracted the image features from four other datasets and used the support vector machine (SVM) for classification. Their research was reached 99.68$\%$ accuracy in the problem of two classes. Nasiri and Hassani \cite{nasiri2022automated} used the DenseNet169 deep neural network to extract image features and the XGBoost algorithm to classify them. They obtained an average accuracy of 98.24$\%$ for the two classes and 98.70$\%$ for the three classes. In 2022, Hasani and Nasiri \cite{hasani2022cov} proposed COV-ADSX, an automated COVID-19 detection system, which used the proposed algorithm in \cite{nasiri2022automated}. They used the Django web framework for the implementation of the COV-ADSX.

Toğaçar\cite{tougaccar2020covid} used X-ray chest images from three classes (COVID-19, Healthy, Pneumonia) and trained the dataset using deep learning models (MobileNetV2, SqueezeNet). Then used SVM for classification. The overall accuracy obtained was 99.27$\%$.
Ucar \cite{ucar2020covidiagnosis} used Bayes deep SqueezeNet to detect COVID-19 based on chest X-ray images and showed an accuracy of 76.37$\%$ for the three classes. Asnaoui\cite{el2021automated} used  X-ray chest images and demonstrated eight transfer learning techniques to classify COVID-19 pneumonia using MobileNet-V2 and Inception-V3 and achieved a 96$\%$ success rate in classification accuracy. Ezzodding et al. \cite{ezzoddin2022diagnosis} employed DenseNet169 and ANOVA feature selection to extract and select features from chest X-ray images. They used LightGBM to classify COVID-19 cases. Thejeshwar\cite{thejeshwar2020precise} used X-ray images of the chest and suggested a KE Sieve Neural Network architecture. The proposed model achieves 98$\%$ accuracy. 
\section{Prerequisites}
The proposed method utilizes DenseNet169 and MobileNet deep neural networks. It also employs LightGBM and the univariate feature selection algorithm. The rest of this section briefly explains these methods.
\subsection{Deep Neural Networks}
The limited amount of data leads to transitional learning to get acceptable results. This research used pre-trained DenseNet169 and MobileNet models to extract image features.
The DenseNet169 network\cite{huang2017densely} is a network of large DenseNet groups already trained in the ImageNet data set. The input of this network is a single image.\\ \indent By connecting each current layer to the previous layers of the network, this network allows the network to be more compact, reducing the number of channels in the network and computational complexity. The DenseNet169 network is 57 MB in size with an accuracy of 93.2$\%$ on the ImageNet dataset.
Google researchers implemented MobileNet Deep Neural Network \cite{howard2017mobilenets} to introduce a lightweight deep neural network. \\ \indent
The Mobile Net Deep Neural Network contains a new type of convolution called depth-wise separable convolution. This new convolution includes a deep convolution filter and a point convolution filter. The deep convolution filter performs a single convolution on each input channel, and the convolution filter combines the deep convolution output points linearly with the 1 × 1 convolution. This network has a size of 16 MB and has reached 89.5$\%$ accuracy on the ImageNet dataset.
\subsection{Univariate Feature Selection}
In univariate feature selection algorithm, $\chi^2$ test is used to find the relation between the variables and the degree of dependency of each independent variable. This test aims to establish the relationship between the two variables. This test distinguishes the degree of independence of the two variables. This test makes it possible to obtain the most useful features in the production of classes.
\subsection{LightGBM}
The LightGBM algorithm was first developed by part of Microsoft's R$\&$D team\cite{ke2017lightgbm}. Although many researchers used XGBoost in their research \cite{akhavan2021internet, chelgani2021interpretable,chelgani2021modeling,nasiri2021prediction,bhattacharyya2022deep,kumar2022classification,fatahi2022modeling}, LightGBM is a more accurate decision tree algorithm than the existing tree boosting algorithms, because LightGBM  produces more complicated trees.
Although producing complex trees, LightGBM is high-speed and uses low memory because it utilizes GOSS and EFB algorithms. Existence of appropriate parameters which allows parameter modification prevents overfitting of the model\cite{ke2017lightgbm}.
\subsection{Gradient-based One-Side Sampling (GOSS)}
In GOSS algorithm the training data are ranked in descending order based on their absolute gradient values. The data with larger gradients have a more significant effect on the data amplification, in which high-gradient samples are sub-sampled by factor a, and low-gradient data samples are sub-sampled by factor b. Therefore, the subset with larger gradient forms subset A and $b*(1-a)$ percent of the data is randomly sampled to form subset B.

By having segmented samples, we use \eqref{eq:1} to evaluate variance gain of data point j with respect to point d.
In \eqref{eq:1}, $x_{i}$ is the $i^{th}$ instance of the training dataset and $g_{i}$ is the negative gradient of the loss function with respect to the $i^{th}$ data. In \eqref{eq:1}, $\frac{1-a}{b}$ is used to normalize the sum of the gradients.
\begin{equation}
\resizebox{0.91\hsize}{!}{$
\bar{V}_{j}(d)=\frac{1}{n}\left(\frac{\left(\sum_{x_{i} \in A_{i}} g_{i}+\frac{1-a}{b} \sum_{x_{i} \in B_{i}} g_{i}\right)^{2}}{n l(d)}+\frac{\left(\sum_{x_{i} \in A_{i}} g_{i}+\frac{1-a}{b} \sum_{x_{i} \in B_{r}} g_{i}\right)^{2}}{n r(d)}\right)$}
\label{eq:1}
\end{equation}
where $ A_{l}$ is the subset of $A$ whose elements are smaller or equal than $d$, and where $ A_{r}$ is the subset of $A$ whose elements are greater than $d$. Similarly $B_l$ and $B_r$ are formed based on set $B$. The definition of $A_l, B_l, A_r$ and $B_r$ is given by:
\begin{equation}
     \begin{aligned}
    A_{l}= \{x_{i}\in A : x_{i,j} \leq d\} , A_{r}=\{x_{i} \in A : x_{i,j} > d\}\\
    B_{l}= \{x_{i}\in B : x_{i,j} \leq d\} , B_{r}=\{x_{i} \in B : x_{i,j} > d\}
    \end{aligned}
    \label{eq:2}
\end{equation}

\subsection{Exclusive Feature Bundling (EFB)}
In deep learning problems, we are confronted with dimensional data; these high dimensions mean many features that reduce the speed of model training. The EFB method reduces the number of features by integrating the data without losing any feature. This process reduces the complexity of the features and speeds up model training.

\section{Proposed Method}
\begin{figure*}[!htbp]
  \centering
  \includegraphics[width=1\linewidth]{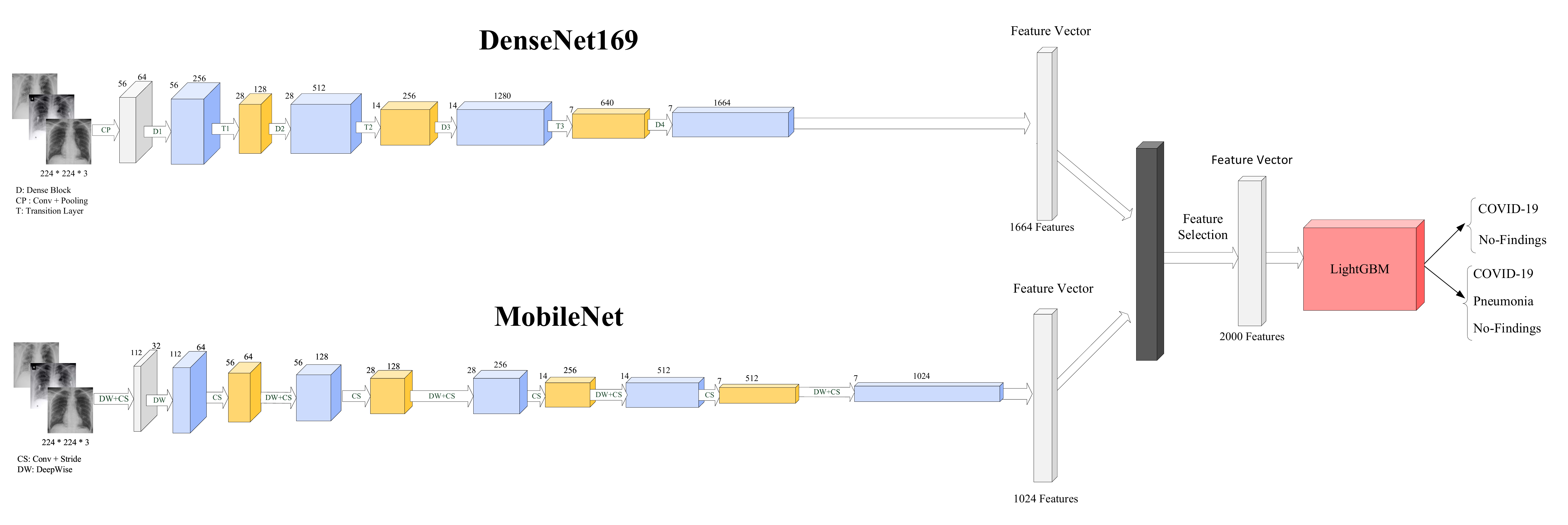}
  \caption{Framework of the proposed method. Pre-trained DenseNet169 and MobileNet extract features separately. Then these features are combined together to increase classification accuracy. Best features are selected to improve the computational complexity of the algorithm. At the end LightGBM is fed with selected features and classifies the new chest X-ray images.}
  \label{Framework}
\end{figure*}

DNNs are very common in feature extraction from medical images \cite{kim2016deep}, therefore we used DNNs in this study for feature extraction from chest X-rays. 
DenseNet169 and MobileNet are the two DNNs that we used for image feature extraction in this research. 
These networks are previously trained by ImageNet database.
Therefore both DenseNet169 and MobileNet can extract features of each new received image. We combined the extracted features from both DNNs to increase the classification accuracy.
Because the number of features increases after the combination, the feature selection algorithm is used to select those features that have the most impact on class differentiation. Finally, the selected features are given to the LightGBM for classification. Fig. \ref{Framework} presents the overall framework for the proposed approach.
\begin{figure}[!ht]
  \centering
  \includegraphics[width=0.9\linewidth]{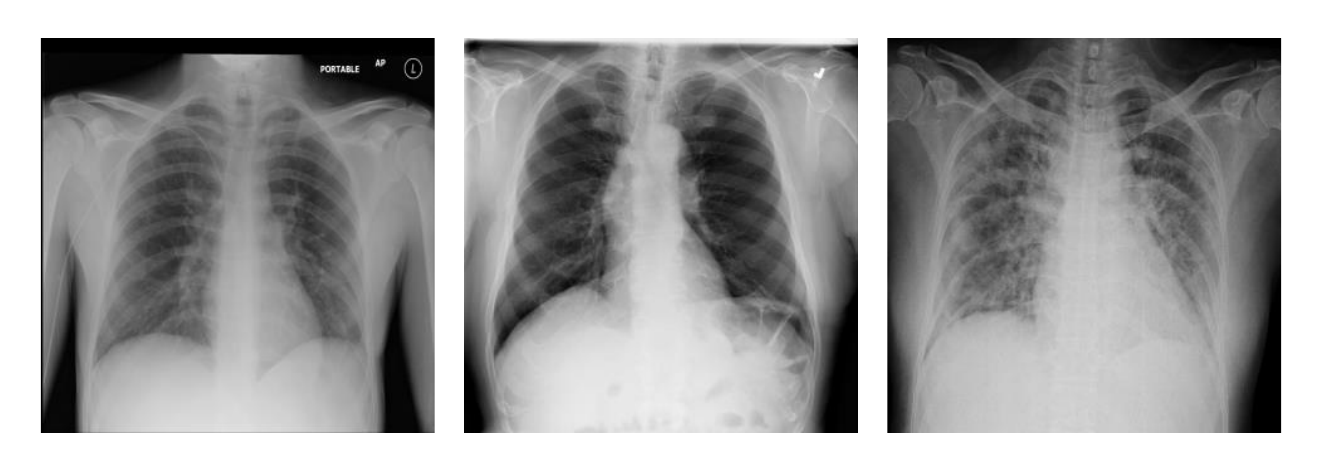}
  \caption{Sample Images in the Dataset}
  \label{sample_images}
\end{figure}
\section{Experimental Results}
To evaluate the proposed method in this paper we used the publicly available ChestX-ray8 image database created by Wang et al. \cite{wang2017chestx}. ChestX-ray8 contains total number of 1125 images, of which 500 are labeled Pneumonia, 500 are labeled healthy, and 125 are labeled COVID-19. 
The total of 1125 images were used to study the problem of the three classes (COVID-19, Pneumonia, Healthy), and 625 images of the dataset were used to study the problem of the two classes (COVID-19, Healthy).\\ \indent
Our research includes three steps. The first step is to extract the features by means of pre-trained DNNs; the second step involves combining the features extracted by DNNs and selecting best extracted features. The third step consists of classifying the relevant images using the LightGBM classification algorithm. A sample of images in the ChestX-ray8 database is shown in Fig. \ref{sample_images}. 
\begin{itemize}
  \item \textbf{Step1:} In this step, 12 pre-trained neural networks were evaluated, and using each of them, image features were extracted and trained for each network of the classifier. Finally, all results have been compared to one another. Due to the accuracy of the LightGBM classifier, the networks that extracted the best features were identified. This study chose DenseNet169 and MobileNet pre-trained neural networks as feature extractors. To extract the features of these neural networks, first, the size of all images has been modified to 244 $\times$ 244. Subsequently, the image's pixel values were normalized and given as input to the deep neural network. Following the operation, the DenseNet169 network extracted 1664 features, while the MobileNet network extracted 1024 features.
  \item \textbf{Step2:} To improve model accuracy, the characteristics extracted by both networks are combined. At this point, a feature vector with 2688 elements is obtained. We know that unnecessary features reduce the efficiency of the model, furthermore, we need to reduce the number of features and extract the most significant features to accelerate the training model.\\
  The univariate feature selection method was used to select the feature, which uses the test $\chi^2$ to choose the best K features. This test computes the degree of independence of each attribute in relation to the classes. Based on the $\chi^2$ test results we chose K to be 2000.
  \item \textbf{Step3:} The input features of the LightGBM algorithm are indicated and the classification is carried out. The reason for choosing this algorithm is due to its high training speed and high accuracy. We selected the LightGBM parameters by trial and error. 
\end{itemize}  

Table \ref{tab:parameters} gives the LightGBM classification parameters. The proposed method for the three-class problem has a 91.11$\%$ accuracy. Furthermore, the average accuracy of the two-class problem with 5-fold cross-validation is 98.54$\%$. Tables \ref{tab2:performance} and \ref{tab1:algorithms} represent the results for problems with three and two classes respectively. The results are compared with the results of Ozturk et al. \cite{ozturk2020automated} and Nasiri and Hassani \cite{nasiri2022automated}. For a better comparison, in addition to accuracy, we also reported sensitivity, specificity, precision and F$_1$ score to achieve a more comprehensive comparison. As shown in Table \ref{tab2:performance}, in the problem of the three classes with respect to the accuracy, the F$_1$ score and the accuracy of our proposed method gives better results than the other methods. With respect to specificity, the method presented by Nasiri and Hassani was the most effective, and our proposed method and the method presented by Nasiri and Hassani performed achieve the highest sensitivity. 

\begin{figure}[!htbp]
  \centering
  \includegraphics[width=0.9\linewidth]{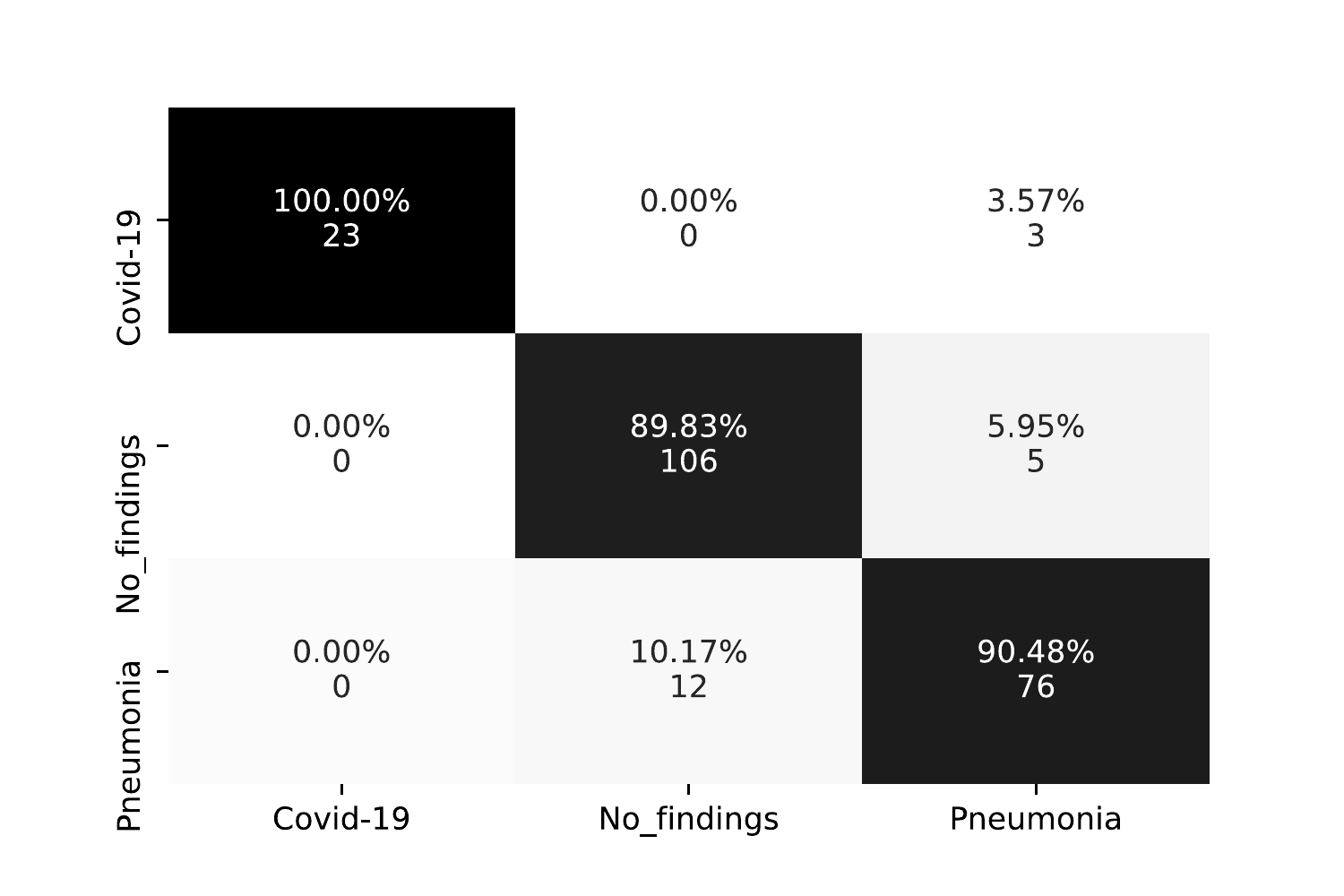}
  \caption{The Confusion Matrix (Three-Class Problem)}
  \label{Multiclass}
\end{figure}

\begin{figure}[!htbp]
	\vspace{-0.72cm}
	\centering
	\begin{subfigure}[b]{0.2\textwidth}
		\centering
		\includegraphics[width=\textwidth]{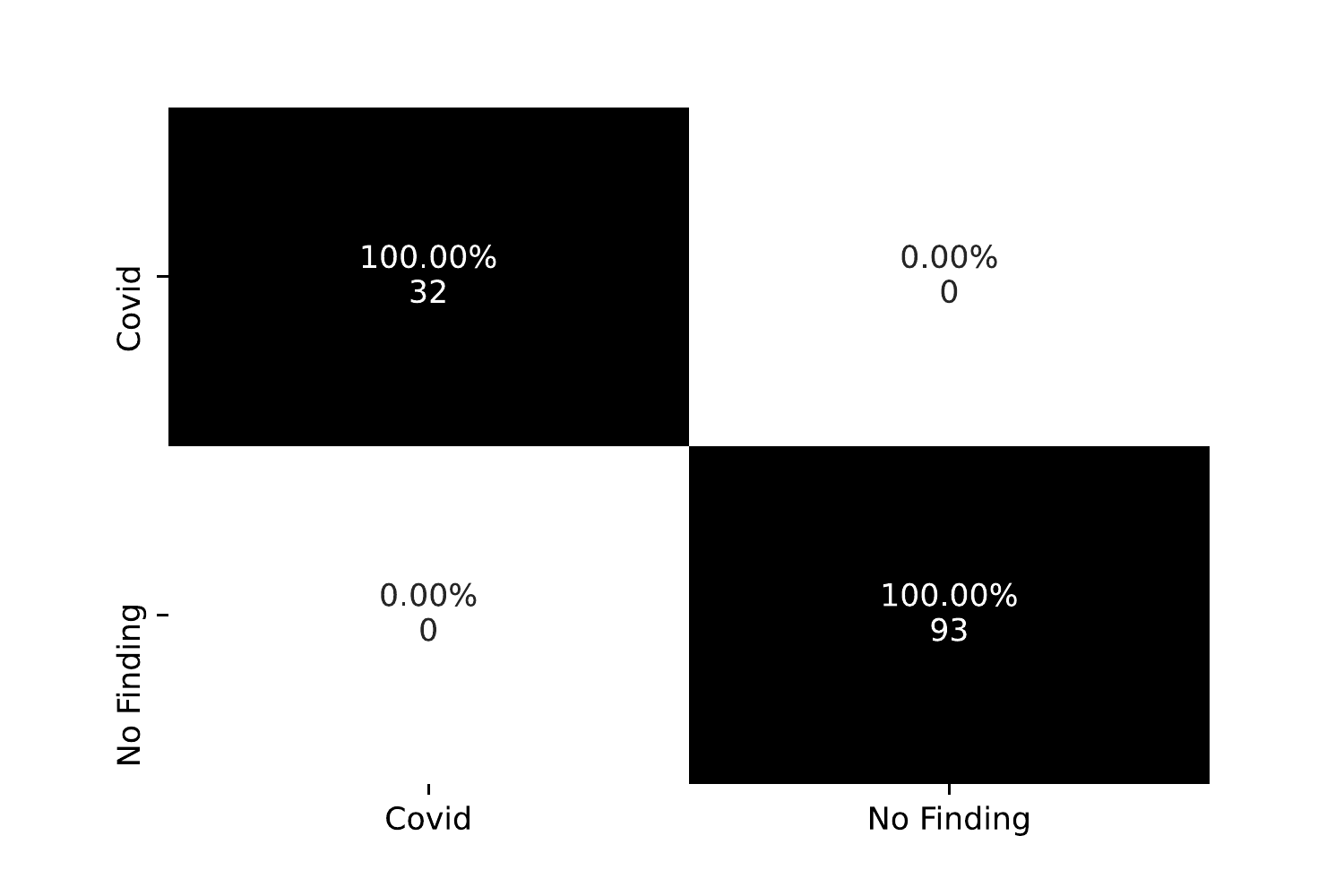}
		\caption{Fold 1}
		\label{fig:Fold1}
	\end{subfigure}
	\hfill
	\begin{subfigure}[b]{0.2\textwidth}
		\centering
		\includegraphics[width=\textwidth]{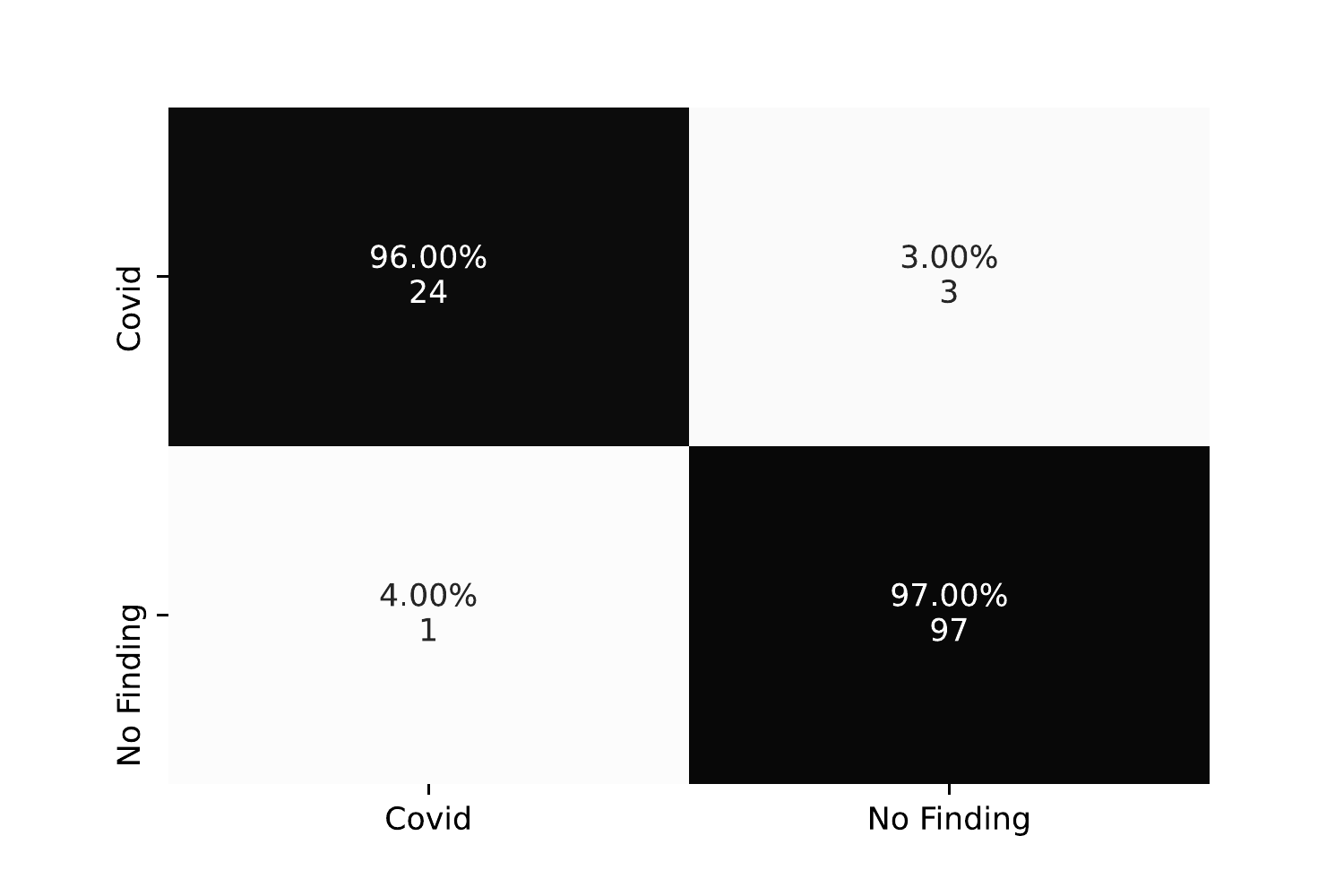}
		\caption{Fold 2}
		\label{fig:Fold2}
	\end{subfigure}
	\hfill
	\begin{subfigure}[b]{0.2\textwidth}
		\centering
		\includegraphics[width=\textwidth]{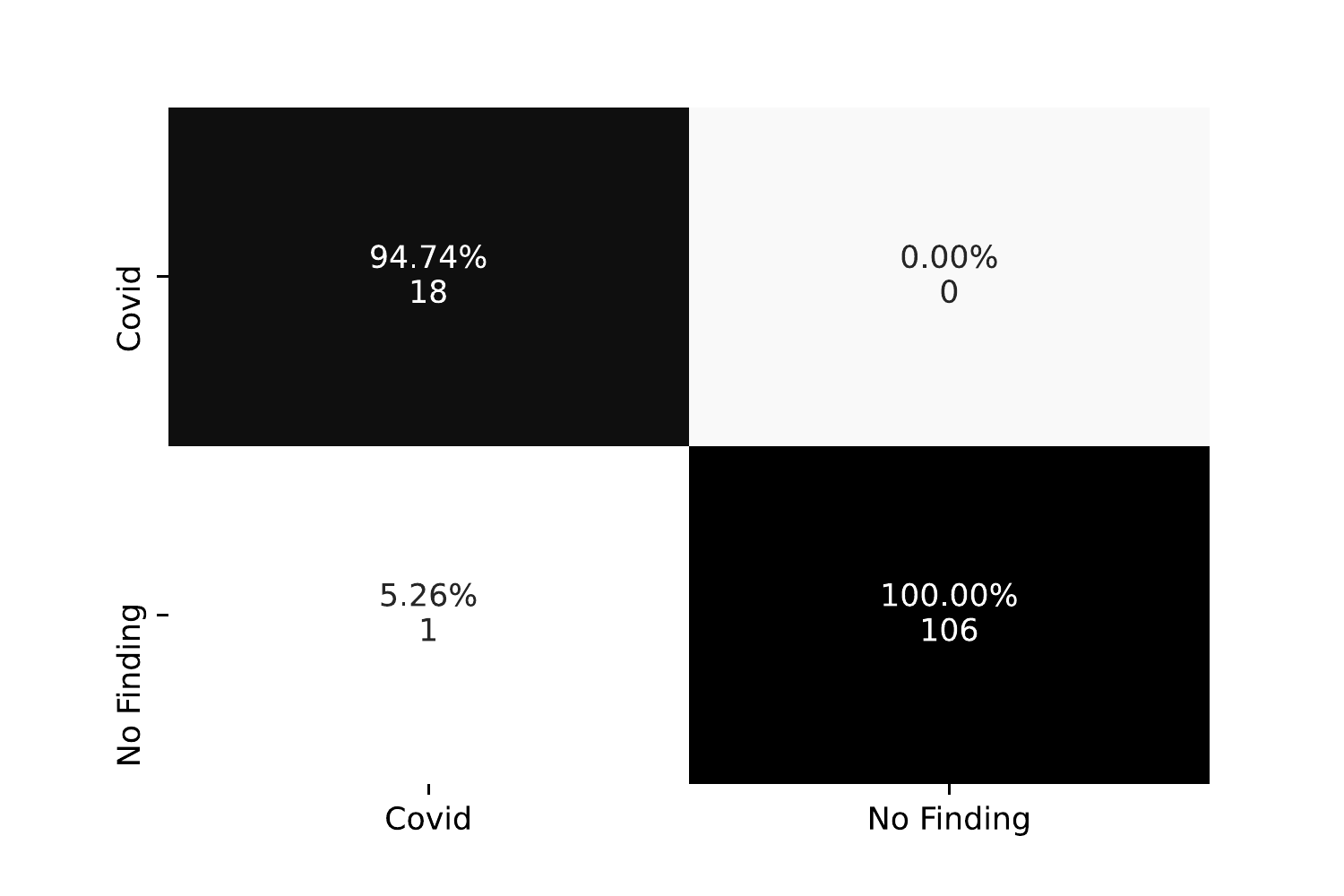}
		\caption{Fold 3}
		\label{fig:Fold3}
	\end{subfigure}
	\hfill
	\begin{subfigure}[b]{0.2\textwidth}
		\centering
		\includegraphics[width=\textwidth]{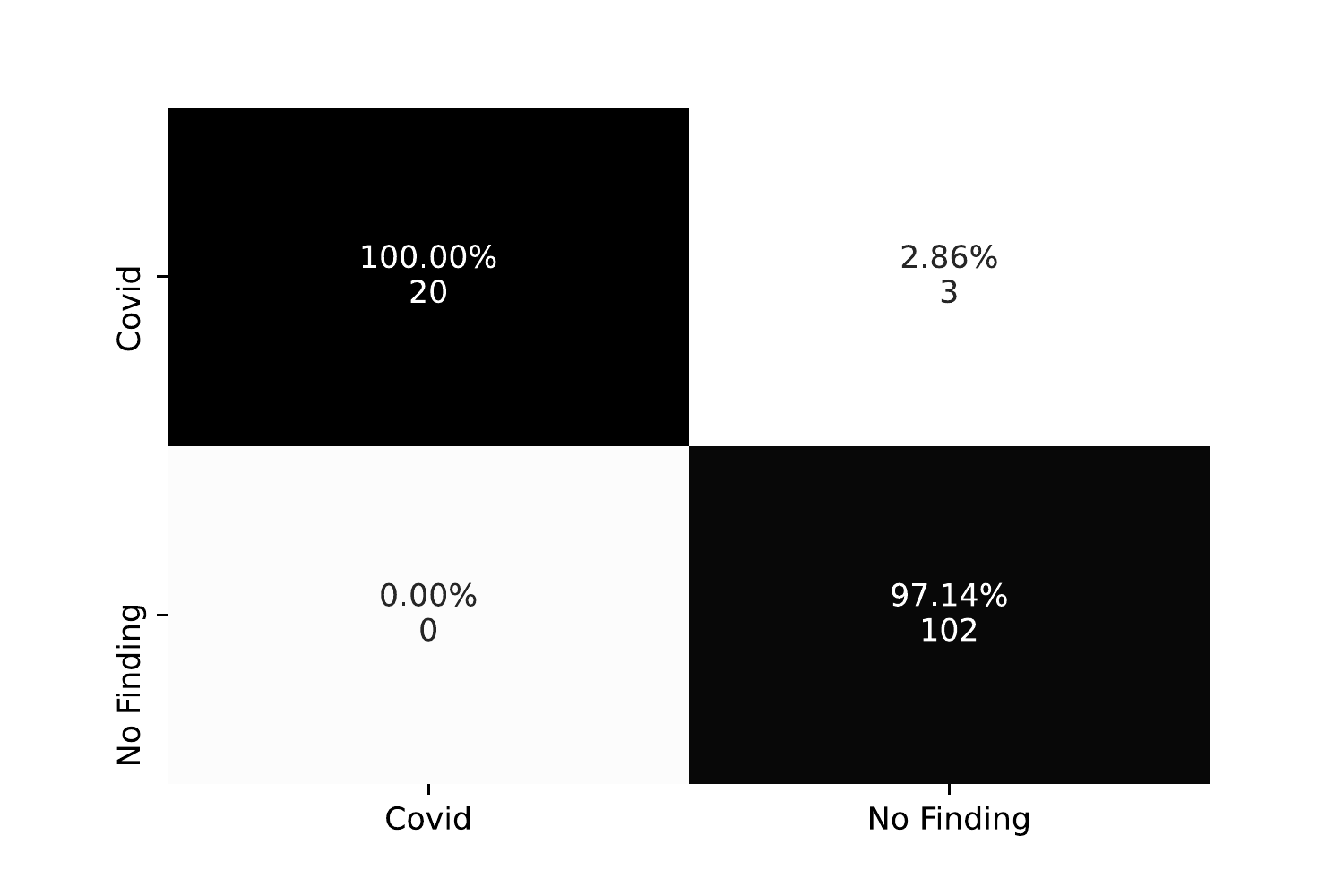}
		\caption{Fold 4}
		\label{fig:Fold4}
	\end{subfigure}
	\hfill
	\begin{subfigure}[b]{0.2\textwidth}
		\centering
		\includegraphics[width=\textwidth]{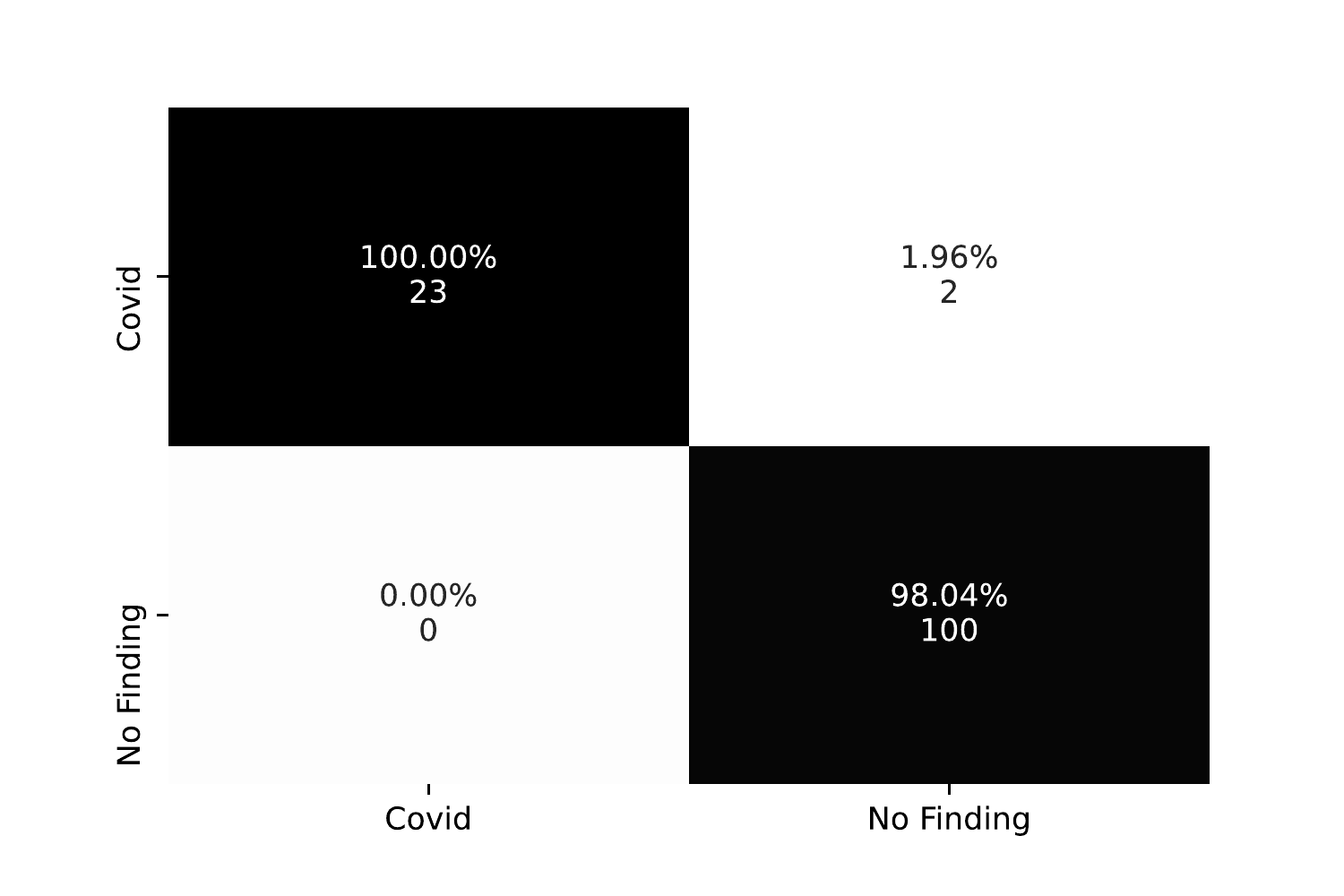}
		\caption{Fold 5}
		\label{fig:Fold5}
	\end{subfigure}
	\hfill
	\begin{subfigure}[b]{0.2\textwidth}
		\centering
		\includegraphics[width=\textwidth]{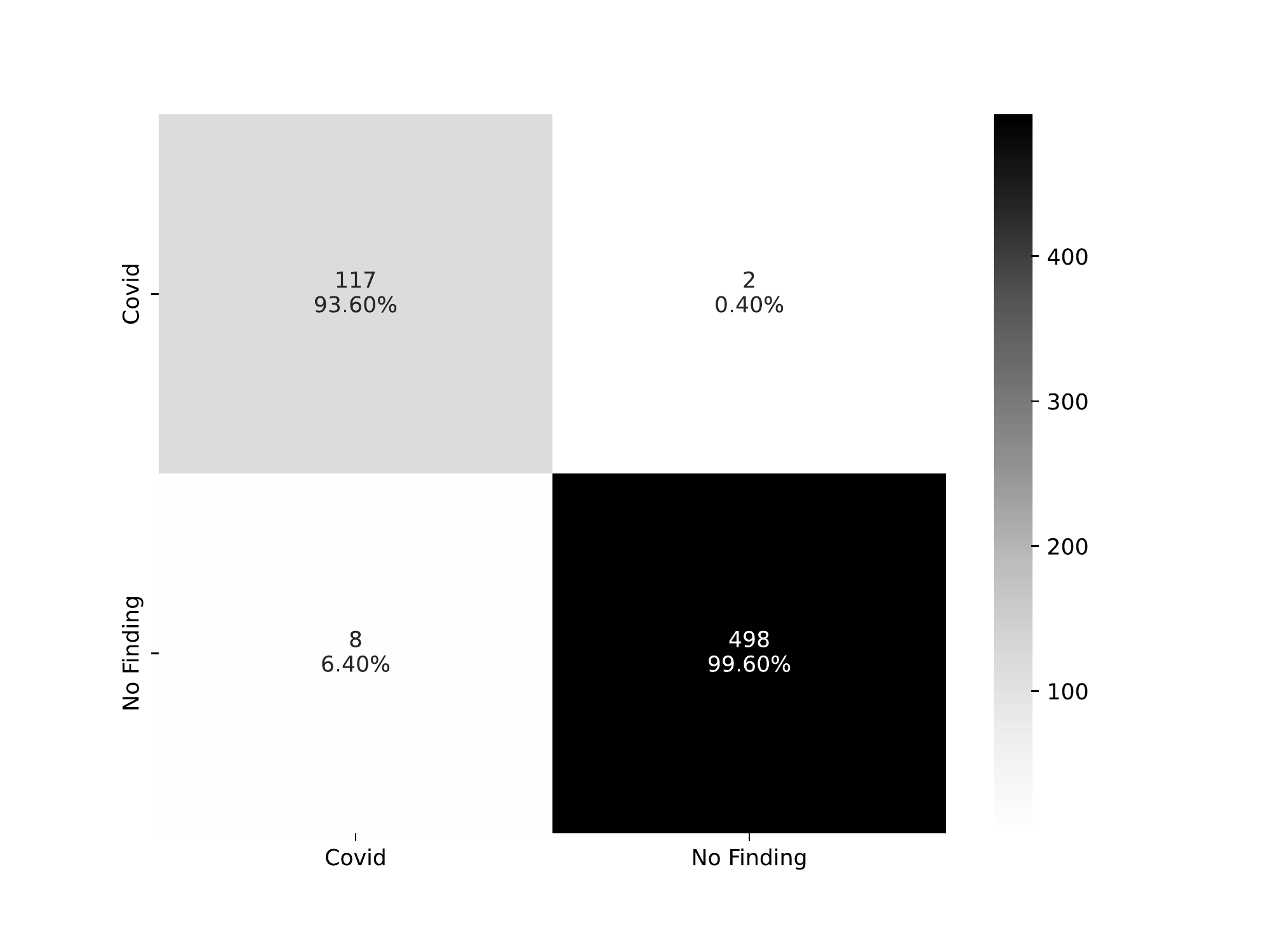}
		\caption{All Data}
		\label{fig:AllFolds}
	\end{subfigure}
	\caption{The Confusion Matrix (Two-Class Problem)}
	\label{fig:ConfusionMatrix_TwoClass}
\end{figure}

In the case of two classes, based on the results in Table \ref{tab1:algorithms}, the sensitivity of our proposed method gives the best results and reaches an average of 99.60$\%$, where the methodology presented by Ozturk et al. is the second. In terms of specificity, the algorithm presented by Nasiri and Hassani outperformes other methods and averaged 99.78$\%$.
Although the results obtained by the three articles are close to each other in terms of precision, but the method presented by Nasiri and Hassani has had the best performance.
With respect to the F$_1$ score and accuracy, the best performance is related to the method proposed in this article.
In both of these criteria, the algorithm presented by Nasiri and Hassani outperformes the method presented by Ozturk et al. It should be noted that our method, due to the use of LightGBM classifier and lack of deep neural network training, has less computational complexity than the method presented by Ozturk et al. Additionally, feature selection and classification were very quickly compared to the Ozturk et al. method. In Figures \ref{Multiclass} and \ref{fig:ConfusionMatrix_TwoClass}, the confusion matrix for three- and two-class problems is shown. Table \ref{tab4:DL-Based} compares the results of the proposed method to other studies. 

\begin{figure}[!htbp]
	\centering
	\includegraphics[width=0.78\linewidth]{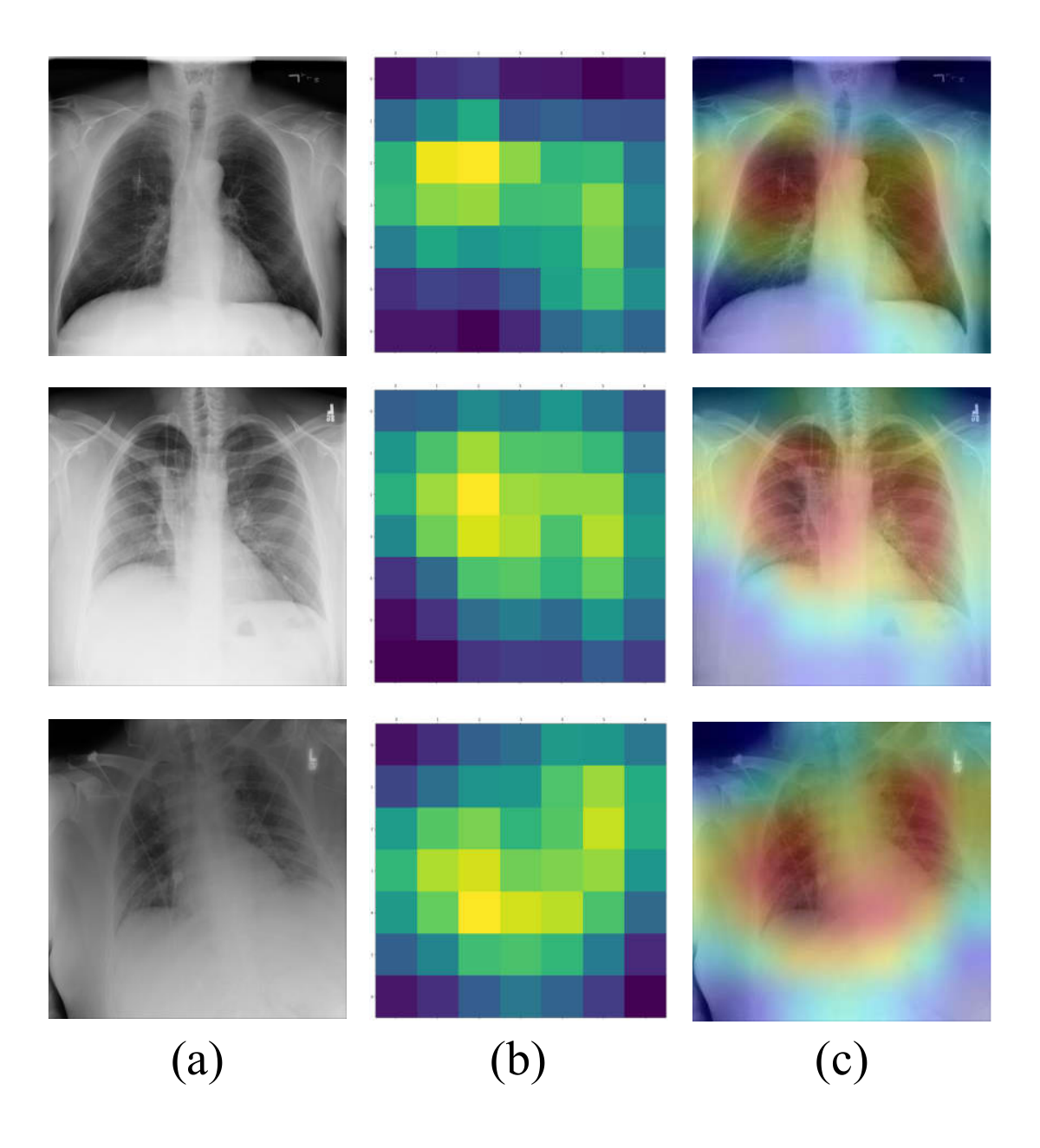}
	\caption{The output of GradCAM for three sample images (a) Original image (b) Class activation heatmap (c) Superimposed image.}
	\label{fig:GradCAM}
\end{figure}

For further assessment of the proposed network, the gradient-based class activation mapping (Grad-CAM) \cite{selvaraju2017grad} was employed to depict the decision region on a heatmap. Fig. \ref{fig:GradCAM} illustrates the class activation heatmaps and superimposed images for three sample images. As can be observed, the proposed method extracted appropriate features, and the model mainly focused on the lung area. Radiologists can use superimposed images to examine the chest area more precisely.

\begin{table}[!htbp]
\renewcommand{\arraystretch}{1.2}
\caption{The LightGBM Parameter Settings}
 
 \centering

 \scalebox{0.8}{

    \begin{tabular}{lc}
    \hline
    Parameter & Value \\
	\hline
	 Learning rate ($\eta$) & 0.24\\
	 Number  of  Trees & 300\\
     Maximum number of leaves   &  105  \\  
     Number of iterations of the algorithm  &  250  \\ 
	 Maximum tree depth &  7  \\ 
	 Minimum number of leaves   & 40  \\ 
	\hline
    	\end{tabular}}
    
	  \label{tab:parameters}
\end{table}

\begin{table*}[t]
\caption{The comparison of the proposed method with other methods (Multi-class problem)}
\centering
\begin{tabular}{p{3cm} p{2cm} p{2cm} p{2cm} p{2cm} p{2cm}}
\hline
\textbf{Method}&\multicolumn{4}{c}{\textbf{Performance}} \\
\cline{2-6} 
 & \textbf{\textit{Sensitivity}}& \textbf{\textit{Specificity}}& \textbf{\textit{Precision}} & 
 \textbf{\textit{F$_1$-score}} & 
\textbf{\textit{Accuracy}} \\
\hline

Proposed Method & \textbf{95.20} & 97.16 & \textbf{95.12} & \textbf{95.6} & \textbf{91.11} \\

\hline
Ozturk et al. & 88.17 & 93.66 & 90.97 & 89.44 & 89.33 \\

\hline
Nasiri \& Hasani & \textbf{95.20} & \textbf{100} & 92.50 & 91.20 & 89.70 \\

\hline
\end{tabular}

\label{tab2:performance}
\end{table*}

\begin{table*}[!htbp]
\caption{Comparison of the proposed method with other methods (Two-class problem)}
\centering
\begin{tabular}{p{3cm} p{2cm} p{1cm} p{1cm} p{1cm} p{1cm} p{1cm} p{1cm}}
\hline
\textbf{Performance}&\multicolumn{7}{c}{\textbf{Different-Folds}} \\
\cline{2-8} 
 & \textbf{\textit{Methods}}& \textbf{\textit{Fold 1}}& \textbf{\textit{Fold 2}} & 
 \textbf{\textit{Fold 3}} & 
\textbf{\textit{Fold 4}} &
\textbf{\textit{Fold 5}} &
\textbf{\textit{Average}}\\
\hline

\multirow{3}{*}{Sensitivity} & Proposed Method & 100 &98.97& 99.06 & 100 & 100 & \textbf{99.60}  \\
 & Ozturk et al. & 100 & 96.42 & 90.47 & 93.75 & 93.18 & 95.13 \\
 & Nasiri \& Hasani & 95.20 & 95.40 & 96.70 & 81.40 & 91.40 & 92.08 \\ 

\hline
\multirow{3}{*}{Specificity} & Proposed Method & 100 & 88.88 & 100 & 86.95 & 92.00 & 93.56  \\
 & Ozturk et al. &  100 & 96.42 & 90.48 & 93.75 & 93.18 & 95.30\\
 & Nasiri \& Hasani & 100 & 100 & 100& 89.90 & 100 & \textbf{99.78} \\

\hline

\multirow{3}{*}{Precision} & Proposed Method & 100 & 97.00 & 100 &97.14 & 98.03 & 98.43  \\
 & Ozturk et al. & 100 & 94.52 & 97.06 & 99.02 & 99.48 & 98.03  \\
 & Nasiri \& Hasani & 99.50& 99.50 & 99.40 & 95.30 & 99.02& \textbf{98.54}  \\

\hline
\multirow{3}{*}{F$_1$-score} &
 Proposed Method& 100 & 97.97 & 97.53 & 98.55 & 99.00 & \textbf{99.01} \\
 & Ozturk et al. & 100 & 98.52 & 93.79 & 95.93 & 95.62 & 96.51   \\
 & Nasiri \& Hasani & 98.50 & 98.50 & 98.20 & 92.50 & 97.30 & 97.00 \\

\hline
\multirow{3}{*}{Accuracy } &
 Proposed Method& 100 & 96.80 & 99.20 & 97.60 & 98.40 & \textbf{98.40}  \\
 & Ozturk et al. & 100 & 97.60 & 96.80 & 97.60 & 97.60 & 98.08  \\
 & Nasiri \& Hasani & 99.20 & 99.20 & 99.20 & 95.20 & 98.40 & 98.24 \\

\hline
\end{tabular}

\label{tab1:algorithms}
\end{table*}

\begin{table*}[htbp]
\caption{Comparison of the proposed method with other DL-Based methods }
\centering
\begin{tabular}{p{3cm} p{3cm} p{5cm} p{3cm} }
\hline

\textbf{Study}&
 \textbf{Number of Samples}& \textbf{Method Used} & 
\textbf{Accuracy} \\
\hline
Apostolopoulos et al.\cite{apostolopoulos2020covid} & 1428 & VGG19 & 93.48  \\
\hline
Wang et al.\cite{gunraj2020covidnet}  & 13645 & COVID-Net & 92.40 \\
\hline
Hemdan et al.\cite{hemdan2020covidx} & 50 & COVIDX-Net & 90.00   \\
\hline
Sethy et al.\cite{sethy2020detection} & 50 & ResNet50+SVM & 95.38 \\
\hline
Narin et al.\cite{narin2021automatic} & 100 & Deep CNN ResNet50 & 98.00  \\
\hline
\multirow{2}{*}{Ozturk et al.\cite{ozturk2020automated}} &
 625& DarkCOVIDNet & 98.08   \\
 & 1125 & DarkCOVIDNet & 89.33  \\
\hline
\multirow{2}{*}{Nasiri \& Hasani\cite{nasiri2022automated}} &
 625& DenseNet169+XGBoost & 98.23  \\
 & 1125 & DenseNet169+XGBoost & 89.70   \\ 
\hline
\multirow{2}{*}{Proposed Method} &
 625& DenseNet169+MobileNet+LightGBM & \textbf{98.54}  \\
 & 1125 & DenseNet169+MobileNet+LightGBM & \textbf{91.11}   \\ 
\hline
\end{tabular}

\label{tab4:DL-Based}
\end{table*}

\section{Conclusion}
This paper presents a novel deep learning based model for the early diagnosis of COVID-19 using the patients chest X-rays. Our proposed method uses DenseNet169 and MobileNet deep neural networks to extract image features, then selects some of the most essential features using the feature selection method, and finally classifies the images using the LightGBM algorithm. Our proposed algorithm achieves 98.54$\%$ accuracy in the classification of two classes of healthy people and COVID-19 patients, and achieves 91.11$\%$ in the classification of three classes (Pneumonia, Healthy, COVID-19).

The evaluation results of our proposed method show that this method has a higher speed and accuracy than previous research in this field due to the use of the LightGBM algorithm.Our novel classification method can help health care providers to diagnose COVID-19 disease which is essential for effective treatment plans. Future work includes verifying the classification results of our method with medical experts. 

\section{Code Availability}
The source code of the proposed method required to reproduce the results is available at the public Github repository\footnote{https://github.com/gha7all/Covid19-Automated-Detection}.

\bibliographystyle{IEEEtran}
\bibliography{IEEEabrv,mybibfile}
\nocite{*}

\end{document}